# Correlations between Background Radiation inside a Multilayer Interleaving Structure, Geomagnetic Activity, and Cosmic Radiation: A Fourth Order Cumulant-based Correlation Analysis

M. E. Iglesias-Martínez[1], J. C. Castro-Palacio[1*], F. Scholkmann[2], V. Milián-Sánchez[3],
P. Fernández de Córdoba[1], A. Mocholí-Salcedo[4], F. Mocholí[4],
V. A. Kolombet[5], V.A. Panchelyuga[5], and G. Verdú[3,6]

[1] Grupo de Modelización Interdisciplinar, InterTech, Instituto Universitario de Matemática Pura y Aplicada, Universitat Politècnica de València, Camino de Vera, s/n, València, Spain

[2] Research Office for Complex Physical and Biological Systems, Zurich, Switzerland

[3] Institute for Industrial, Radiophysical and Environmental Safety, Universitat Politècnica de València, Camino de Vera, s/n, València, Spain

[4] Traffic Control Systems Group, ITACA Institute, Universitat Politècnica de València, Camino de Vera, s/n, València, Spain

[5] Institute of Theoretical and Experimental Biophysics, Russian Academy of Science, Moscow Region, Pushchino, Russia

[6] Chemical and Nuclear Engineering Department, Universitat Politécnica de València, Camino de Vera, s/n, València, Spain

(*) Corresponding author

**Abstract`**

In this work, we analyzed time-series of background radiation inside a multilayer interleaving structure, geomagnetic activity and cosmic-ray activity using the Pearson correlation coefficient and a new correlation measure based on the one-dimensional component of the fourth order cumulant. The new method is proposed based on the fact that the cumulant of a random process is zero if it is of Gaussian nature. The results show that this methodology is useful for detecting correlations between the analyzed variables.

## 1. Introduction

In a previous work our group [1] we showed that the measured radioactive decay rates of different nuclides changed significantly when placed inside a simple enclosure in form of a modified Faraday cage [1, 2]. The variations ranged at least between 0.8% and 5%. This enclosure also showed to cause anomalous capacitance measurements in ultra-stable capacitors, as well as in the measurements obtained with other devices. For example, the time constant of a RC low-pass filter increased at least by 5.5%, and the spectrum of a Cs-137 source was distorted, i.e. the photopeak was shifted to lower energies and its height increased.

In subsequent works [3, 4] we found correlations between some of the anomalous decay processes [1] and geomagnetic activity (GMA) as well as cosmic-ray activity (CRA). This finding is important since it showed that the measured variability in reference [1] was not caused by some failure in the electronics of the instruments but rather followed the variations in GMA and CRA. Besides, it made apparent that a link exists between GMA/CRA and the outputs of the measuring system, i.e., the measured radioactive decay rates and the capacitance values.

In our most recent analysis [4] we analyzed the correlation with linear correlation analysis and statistical testing based on Bayesian statistics. The correlation analysis revealed novel insights with regard to the relationship between the measured radioactive decay and GMA/CRA. We proposed some preliminary conclusions about the conditions under which those correlations took place, and on the circumstances in which the correlations could not be registered. One of the main conclusions was that the correlations between decay activity (counts per minute registered in the Geiger-Muller (GM) counter) and GMA/CRA could take place (at least when working with Ra-226) in those time intervals when the decay values increased (or were increased with regard to the counts outside the box). Although our analysis revealed new insights we recognized that the correlation analysis had limitations in case of comparing time-series with correlations. Therefore, we aimed to extend the correlation analysis with a novel approach. TO this end, correlations are analyzed this time from different points of view, that is, using fourth order statistics as by this means one can draw more conclusions about the possible existing correlations. In the present paper only the background radiation measured with the GM counter was studied. In upcoming papers, the other processes described in [1] (i.e. decay rates with some radionuclides and capacitance variations) will be analyzed with the new approach as well.

## 2. Experimental setup and data analysis

The key elements of the setup were a Faraday-like cage (developed by Reich [5]), whose sides are formed by interleaving sheets of metallic and organic materials (Fig. 1a), and the Geiger-Müller counter tube (Fig. 1b and Fig. 1c):

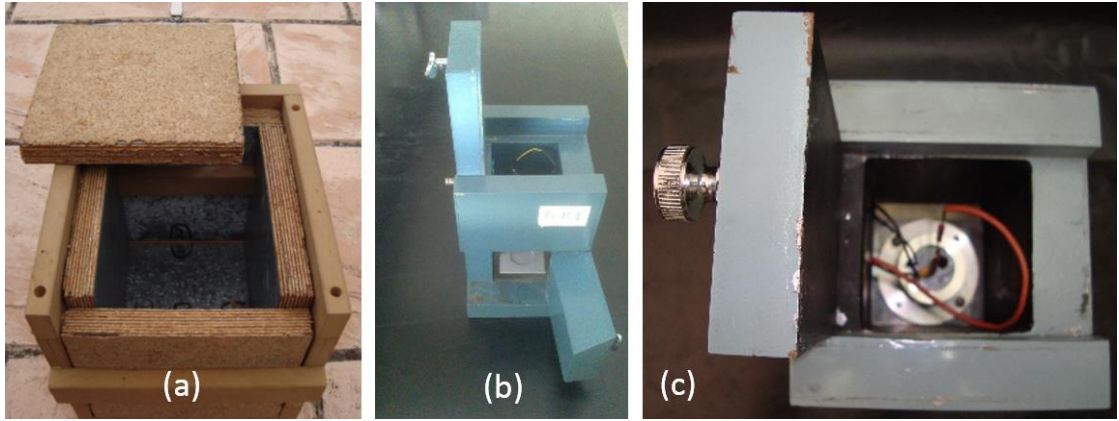

Fig. 1. (a) A small model of metallic shield in form of Faraday cage where the sides are formed by interleaving structures of metallic and organic (cork sheeting). (b) Lead shield that contains the Geiger-Müller counter tube. The tube is inserted inside the shield on the upper part opening as looks as in (c).

On the other side, in the development of the work, an algorithm was proposed as a measure of similarity and features extraction, which it is based on a time-correlation using the Pearson correlation coefficient of the one-dimensional component of fourth order cumulant calculation.

### 2.1. Linear correlation analysis

As a first analysis and to determine whether or not there is a correlation in the data, Pearson's correlation coefficient was used as a measure to assess the similarity of the data in a time series, based on the covariance matrix.

Let $Y = \{Y(t)\}_{t \geq 0}$ be a zeroth-mean stochastic process, then the autocovariance is a measure of the dispersion of the process about the mean value, and can be defined in terms of its joint moments, see for instance references [6-9], where $X(t)$ is the determinist part (i.e. the signal to be detected) and $n(t)$ is a random process considered as additive noise.

$$C_2^Y(\tau) = E[(X(t) \cdot X(t+\tau))] + E[(n(t) \cdot n(t+\tau))], \qquad (1)$$

where $C_2^Y(\tau)$ stands for the second order cumulant. For the discrete model, data equation (1) becomes,

$$C_2^Y(\tau) = \sum_{t=0}^{M-1-\tau} X(t) \cdot X(t+\tau) + \frac{1}{N} \sum_{t=0}^{M-1-\tau} n(t) \cdot n(t+\tau), \qquad (2)$$

where $M$ is the number of data points to be processed. The correlation coefficient is then given based on the covariance function:

$$R(i,j) = \frac{C_2(x,y)}{\sqrt{C_2(x,x) C_2(y,y)}}, \qquad (3)$$

where $R(i,j)$ is the correlation matrix and $C_2(i,j)$ is the covariance matrix of the two data set to be analyzed.

The use of the covariance matrix to obtain the Pearson correlation coefficient has the disadvantage of not being immune to noise. This means that if the data to be analysed presents noise, it can influence the correlation levels obtained, of one variable with respect to the other.

From previous discussion, it is proposed to use higher order statistics for the calculation of the correlation matrix. To this end, the use of the one-dimensional component of the fourth order cumulant of the data was suggested, which for Gaussian processes becomes zero.

### 2.2. Algorithm Based on Second Order Statistics: Fourth Order Cumulant-based Correlation Analysis

For zero-average Gaussian processes, the fourth order cumulant can be calculated as[4]

$$c_4^Y(\tau_1,\tau_2,\tau_3) = E\{Y(t) \cdot Y(t+\tau_1) \cdot Y(t+\tau_2) \cdot Y(t+\tau_3)\} \\ -c_2^Y(\tau_1) \cdot c_2^Y(\tau_2-\tau_3) - c_2^Y(\tau_2) \cdot c_2^Y(\tau_3-\tau_1) - c_2^Y(\tau_3) \cdot c_2^Y(\tau_1-\tau_2) \quad (4)$$

Working with the one-dimensional component of the fourth order cumulant of the signal, $c_4^Y(\tau_1,0,0)$, and when setting $\tau_2 = \tau_3 = 0$, the same result as in [6] is obtained (it was obtained very similarly, but doing $\tau_1 = \tau_2 = \tau_3 = \tau$). We obtain the following:

$$c_4^Y(\tau_1,0,0) = E\{Y(t)^3 \cdot Y(t+\tau_1)\} - 3 \cdot E\{Y(t) \cdot Y(t+\tau_1)\} \cdot E\{Y^2(t)\} \quad (5)$$

For a discrete number of data points, Eq. (5) can be rewritten as,

$$C_4^Y(\tau) = \sum_{t=0}^{M-1-\tau} Y^3(t) \cdot Y(t+\tau) - 3\left\{\frac{1}{N}\sum_{t=0}^{M-1-\tau} Y(t) \cdot Y(t+\tau)\right\} \cdot \left\{\sum_{t=0}^{M-1} Y^2(t)\right\} \quad (6)$$

Thus, we can evaluate the correlation matrix,

$$R(i,j) = \frac{C_4(i,j)}{\sqrt{C_4(i,i)C_4(j,j)}} \quad (7)$$

where $R(i,j)$ is the correlation matrix using higher order statistics, and $C_4(i,j)$ is the one-dimensional component of the fourth order cumulant.

## 3. Application results

In order to check the analysis approaches described above, the data of one of the tests performed inside the cage was analysed. This test (denoted by D1) yields a positive result, thus all the other tests about background counts measured inside the box were also analysed in the same way. All those results are presented in what follows.

In the aforementioned test, the data were obtained by measurements inside the box starting on 2014 December 19 at 17:00 hours and finishing on 2014 December 22 at 09:00 hours. Those data are presented in [2] and the results of the correlation analysis are shown in **Table 1**. The total number of processed samples was 65.

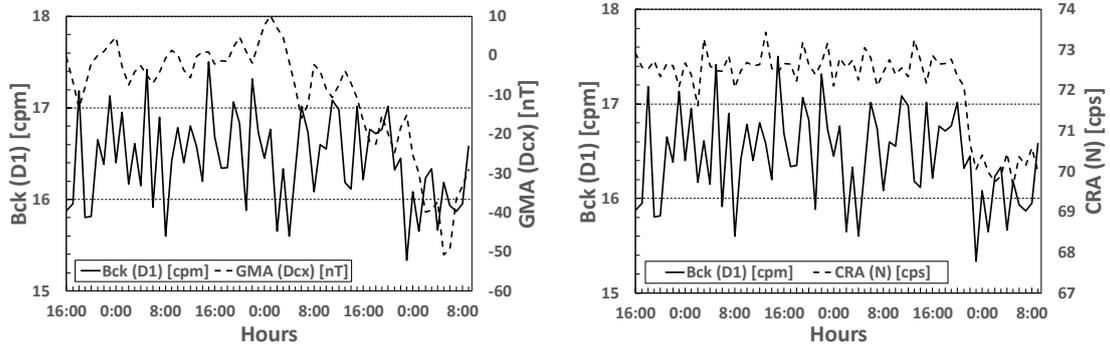

Fig. 2 Background data (D1), GMA (Dcx) and CRA (N) between 2014 December 19 and 2014 December 22. Dcx: Dcx index. N: neutron counts.

**Table 1** shows that in this specific case (where covariance is used), there is a high correlation between the background (Bck) fluctuations (D1) and CRA and the GMA with CRA, respectively.

*Table 1*
*Correlation between the three variables background (D1),*
*GMA (Dcx) and CRA (N), using covariance (equation 3)*

| **Correlations between variables** | | |
|---|---|---|
| **Bck-GMA** | **Bck-CRA** | **GMA-CRA** |
| 0.2441 | 0.4104 | 0.7927 |

The above is consistent with the work done in [4], whose results are based on the correlation using the Spearman coefficient and show that there is a small correlation between the background and GMA (0.2441) and a clear correlation between the background and CRA (0.4104).

Both algorithms, the first one used in this work (Eq. 3) as well as the one used in [4], are based on the covariance matrix of the data. Spearman's correlation coefficient (used in [4]) is less sensitive than Pearson's coefficient (used in the present paper) for values that are far from expected, that is, for isolated values that do not predominate in a set. To illustrate these results and to make a comparison with those obtained in [4], see table 2.

*Table 2*
*Correlation between the three variables background (D1),*
*GMA (Dcx) and CRA (N). A comparative study using*
*Pearson and Spearman Coefficients*

| Coefficient | Correlations between variables | | |
|---|---|---|---|
| | **Bck-GMA** | **Bck-CRA** | **GMA-CRA** |
| Pearson | 0.2441 | 0.4104 | 0.7927 |
| Spearman | 0.1691 | 0.3528 | 0.4594 |

After analysing the correlations pairs using the one-dimensional component of the fourth order cumulant, the following results were obtained (**Table 33**):

*Table 3*
*Correlation between the three variables background (D1),*
*GMA (Dcx) and CRA (N), using the one-dimensional component of*
*fourth order cumulant (equation 7)*

| Correlations between variables | | |
|---|---|---|
| **Bck-GMA** | **Bck-CRA** | **GMA-CRA** |
| -0.1972 | -0.4432 | 0.6751 |

**Table 3** shows that there is a linear correlation between the three analyzed data. In two of the cases the results show a negative correlation. All the other available measurements of the background inside the cage were analyzed with this methodology. The results are presented in **Table 4**.

Table corresponds to the period labelled as J1 (2014 July, 10). From the experimental data measured on that day, taken from 08 hours until 12 hours, there is a high correlation between the three variables, using both covariance (second order statistics) and cumulant (four order statistics).

The second analysis in **Table 4** corresponds to the period labelled as J2 (2014 July, 11). From the obtained experimental data, taken from 09 hours until 11 hours, one can evaluate the correlation between the variables using covariance (second order statistics) and cumulant (four order statistics).

The last analysis in **Table 4** corresponds to the period labelled as D1 (2014 December, 19). The experimental data were measured, starting on 2014 December 19, (at 17:00 hours) and finishing on 2014 December 22, 2014 (at 09:00 hours).

**Table 4** shows that when there is a correlation between the environmental variables (geomagnetic activity and cosmic neutrons flux), there is a correlation between the background and these variables.

*Table 4*
*Correlation between the three variables background (D1), GMA (Dcx) and CRA (N).*

| Data | Using covariance (Eq.3) | | | Using the one-dimensional component of fourth order cumulant (Eq.7) | | |
| --- | --- | --- | --- | --- | --- | --- |
| | **Bck-GMA** | **Bck-CRA** | **GMA-CRA** | **Bck-GMA** | **Bck-CRA** | **GMA-CRA** |
| J1 | 0.8735 | 0.8747 | 1.000 | 0.9401 | 0.9407 | 1.000 |
| J2 | 0.1166 | -0.0001 | 0.0657 | -0.2628 | -0.0156 | 0.123 |
| D1 | 0.2448 | 0.4104 | 0.7948 | -0.1974 | -0.4432 | 0.6705 |

Specifically, when there is a correlation above 0.6 between GMA and CRA, there is a non-negligible probability of finding a correlation between the background radiation and these variables.

The cases just shown analysed few experimental data. Thus, more measurements are necessary to gain more insights into their behavior under similar experimental conditions and to understand the involved phenomena.

**Conclusions**

In this work, two methods to study the correlations have been shown, one based on using the Pearson correlation coefficient as the first measure of similarity, and another using a correlation measure based on the one-dimensional component of the fourth order cumulant.

From the experimental results and from both methods of analysis, we can conclude that:
(i) There is a linear, direct (as well as inverse) relationship between the background-GMA and background-CRA data when there is a correlation higher than 0.6 between the space weather variables, Dcx and *N*.
(ii) We have carried out the experimental tests in different periods (called J1, J2 and D1) and our methodology reveals the presence (or not) of correlations in those periods. Due to these facts, we consider that it is necessary to establish a more complete measurement planning in order to derive a model to understand this phenomenon.

**Author Contributions:** All authors contributed equally.

**Funding:** This research was supported by grant no. RTI2018-102256-B-I00 (Spain).

**Conflicts of Interest**: The authors declare no conflict of interest.